\documentclass{epl}

\title{Statistics of precursors to fingering processes}
\shorttitle{Statistics of precursors}
\author{Patrick Grosfils \inst{1} \and Jean Pierre Boon \inst{1}} 
\institute{
  \inst{1} Physics Department CP 231,
       Universit\'e Libre de Bruxelles, \\
       1050 - Bruxelles, Belgium
       (E-mail: {\tt pgrosfi@ulb.ac.be} ; {\tt jpboon@ulb.ac.be})
}

\newcommand{\marge}[1]{\marginpar{}}  
\newcommand{\Sl}[1]{{}}           
\newcommand{\beq}[1]{\Sl{#1}\begin{equation}\if#1\empty\else\label{#1}\fi}
\newcommand{\eeq}{\end{equation}}
\newcommand{\beqa}[1]{\Sl{#1}\begin{eqnarray}\if#1\empty\else\label{#1}\fi}
\newcommand{\eeqa}{\end{eqnarray}}

\newcommand{\Eq}[1]{(\ref{#1})}

\newcommand{\ov}{\overline}

\begin{document}

\maketitle

\begin{abstract}
We present an analysis of the statistical properties of hydrodynamic field 
fluctuations which reveal the existence of precursors to fingering processes. 
These precursors are found to exhibit power law distributions, and these power 
laws are shown to follow from spatial $q$-Gaussian structures which are 
solutions to the generalized non-linear diffusion equation.
\end{abstract}

Over the past fifteen years, a nonextensive variant to the Boltzmann-Gibbs 
formulation of Statistical Mechanics has been developed for the analysis of 
systems away from the equilibrium state \cite{tsallis} characterized by the
appearance of non-exponential distributions and power laws. 
For instance, for diffusion type processes, non-exponential distributions are 
obtained from generalized Fokker-Planck type equations, and it has been 
shown, in a generic manner from a generalization of classical linear response 
theory, that such distributions may arise from first-principle 
considerations~\cite{lutsko-boon}. The goal of this letter is (i) to show that 
in fingering phenomena, patterns are preceded by precursor processes, 
and (ii) to present an analysis of the statistical properties of hydrodynamic field fluctuations in these precursors; we find power law distributions which are 
shown to follow from the solution of the generalized non-linear diffusion equation.
Although we do not use nonextensive statistical mechanics to {\em explain} our 
results, we find mathematical similarities which suggest a possible connection
as discussed in the concluding paragraph. 

The phenomena investigated here arise {\em before} the onset 
of fingering, a generic phenomenon that results from the destabilization of the 
interface between two fluids with different mobilities in systems such as a shallow 
layer or a porous medium, when the fluid with highest mobility is forced 
through the medium filled with the other fluid. Here {\em before} means
that the constrained fluid is in a state where no fingering pattern is as yet visible, 
but where hydrodynamic field fluctuations are enhanced as {\em precursors}
to the onset of fingering. The analytical form of the statistical properties of 
these precursors are compatible with the solution of the generalized diffusion 
equation~\cite{lutsko-boon} which has formally the same structure as the
``porous media equation" \cite{compte}, but where the diffusion coefficient depends 
on the solution of the equation. This leads to the fact that the diffusion process 
is classical in the sense that there is linear scaling with time, but the solutions 
are not Gaussian: they have the canonical $q$-exponential form \cite{tsallis}.

To date - at least to the best of our knowledge - such precursor properties 
have not yet been obtained from laboratory measurements. 
Here we use two methods: (1) a mesoscopic approach, the lattice Boltzmann 
simulation method \cite{succi}, which is based on a kinetic theoretical analysis 
where the macroscopic description is not pre-established, and (2) a 
phenomenological approach, Darcy's law \cite{darcy} which is solved
numerically. Our results show that (i) the system's non-linear response to the 
driving force produces dynamical behavior in accord with the generalized diffusion 
equation while no such phenomenological prescription was injected a priori, 
(ii) the statistics of the hydrodynamic fluctuations reveal the existence of
precursor processes before fingering patterns become visible, 
(iii) the distribution of these field fluctuations are given by power laws 
(with an exponent $|q|<1$), and 
(iv) these power laws follow from space integration of two-dimensional
$q$-Gaussian  structures.

These structures form a landscape of alternating upward and downward 
$q$-Gaussian ``blobs" (see Fig.2 below) which are precursors to the
fingers, and are reminiscent of vortices with alternating parities in 
2-$d$ turbulence 
\footnote{Here the Reynolds number has low value, but 
the relevant control parameter, the P\'eclet number, is high.}.
For instance, experiments in turbulent Couette-Taylor flow \cite{swinney} 
have shown that data obtained from quantities extracted from velocity field measurements exhibit  $q$-exponential distributions which have been analyzed 
with theoretical arguments based on nonextensive statistical mechanics \cite{beck}. Other physical systems have been shown to exhibit similar type distributions 
interpreted as a consequence of {\it nonextensivity}  \cite{tsallis} or 
{\it superstatistics} \cite{beck-cohen}. 
We suggest that in the early stage of the fingering process, the signature 
of nonextensivity can be found in the recasting of interacting Gaussian 
structures into a sum of independent $q$-Gaussian blobs.  

\begin{figure}
\begin{center}
\makebox{
\rotatebox{-90}{
\resizebox{6.7cm}{8cm}
{\includegraphics{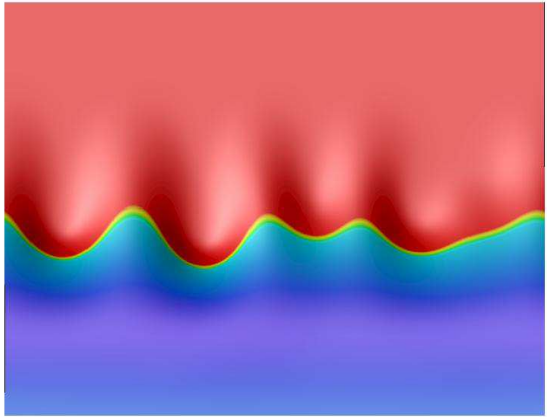}}}}
\caption{ Illustration of lattice Boltzmann (LBGK) simulation of 
              viscous fingering (flow direction downwards).
               The mixing length between the two fluids is clearly identified by the 
                shadow zone around the concentration $=0.5$ contour line
               (color code indicating the concentration 
                of the invading fluid from red ($C_1=1$) to blue ($C_1=0$)).}
\end{center}  
\label{Fig_fingering}
\end{figure} 

We consider a two miscible fluids system confined between two plates 
with narrow gap, a configuration known as the Hele-Shaw cell \cite{homsy}. 
The Hele-Shaw geometry is intrinsically three-dimensional, but the effective destabilization of the flow can be described in the 2-$d$ plane, given that the 
flow has a Poiseuille profile in the third dimension. So Hele-Shaw flow can be 
simulated with the two-dimensional LBGK equation, the 
Bhatnagar-Gross-Krook single relaxation version of the lattice Boltzmann 
equation \cite{succi}. One emulates 3-$d$ flow by introducing a drag term,
thereby simulating a system with a virtual cell gap in the 
third dimension: the drag enters the LBGK equation as a damping term.
The simulation method has been described elsewhere
\cite{boek}: essentially the system consists of a 2-$d$
box with vertical length $L_y = 2048$ nodes, and horizontal width 
$L_x = 1024$ nodes, with horizontal periodic boundary conditions; 
initially the box is filled with fluid 2 which is 
then displaced by fluid 1 injected uniformly from the top of the box.
Fluid 1 (with damping coefficient $\beta_1 < \beta_2$) invades 
the system through a constant force applied along the $y$-direction, 
which produces a flow with P\'eclet number $Pe \simeq 200$.

The initial concentration profile is a step function, which because of 
mutual diffusion, develops into an erfc~$y$, on top of which  white noise 
is imposed (along the $x$-direction) to trigger the instability.
When the instability develops, it produces a moving fingering pattern 
as illustrated in Fig.1. Here we consider the early stage where fingers are
not yet visible (they have not developed as in Fig.1 shown here for the 
purpose of illustration), just below the dynamical transition where the 
exponent $\mu$ of the growth of the mixing length of the interfacial zone, $L_{mix}\propto t^{\mu}$, changes from $\mu = 1/2$ (diffusive regime) 
to $\mu \simeq 2.3$ (see Fig.3 in \cite{boek}). In this early time regime, 
the flow produces local concentration gradients which induce 
mobility fluctuations thereby triggering vorticity fluctuations.

\begin{figure}
\begin{center}
\makebox{
\rotatebox{-90}{
\resizebox{6cm}{8cm}{
\includegraphics{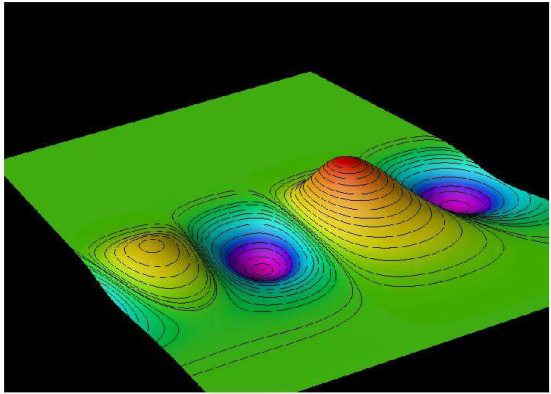}}}}
\caption{Concentration fluctuations field showing landscape of $q$-Gaussian 
``hills and wells" (flow direction South-East); simulation data with concentration  
contour lines, color code indicating highest positive values (red) to largest 
negative values (magenta) of~$c$.}
\end{center}  
\label{Fig_C_landscape}
\end{figure} 

We measured the following local fluctuations:\\
$\cdot$ concentration field (fluid 1): $c(x,y) = C(x,y)-\ov{C(x,y)}$;\\
$\cdot$ transverse velocity field: $v_x(x,y) = V_x(x,y)-\ov{V_x(x,y)}$;\\
$\cdot$ longitudinal velocity field: $v_y(x,y) = V_y(x,y)-\ov{V_y(x,y)}$;\\
$\cdot$  and vorticity field: $\omega(x,y) = \partial_x v_y(x,y) - \partial_y v_x(x,y)$,\\
where the overline denotes average along the $x$-direction.
The concentration field in Fig.2 shows a ``landscape" of alternating hills and wells. 
In each blob, the concentration field exhibits a two-dimensional $q$-Gaussian 
profile as illustrated in the left panel of Fig.3 obtained by a section plane cut 
through the extrema in Fig.2 parallel to the $x$-axis.  
The corresponding concentration distribution, $P(c)$, is obtained by taking 
the values of $c(x,y)$ measured over all space, and is shown to follow a power law, 
$P(c) \propto |c|^{-q}$ with $q=0.73$ (see right panel of Fig.3). Note that the power 
law exponent stays the same over the entire range of the distribution.
The other local fields exhibit similar behavior and the corresponding 
distributions $P(v_x), P(v_y)$ and $P(\omega)$ all show power laws 
(with an exponent $<1$) which follow from space integration (see below).

\begin{figure}
\begin{center}
\makebox{
\rotatebox{-90}{
\resizebox{6cm}{7cm}{
\includegraphics{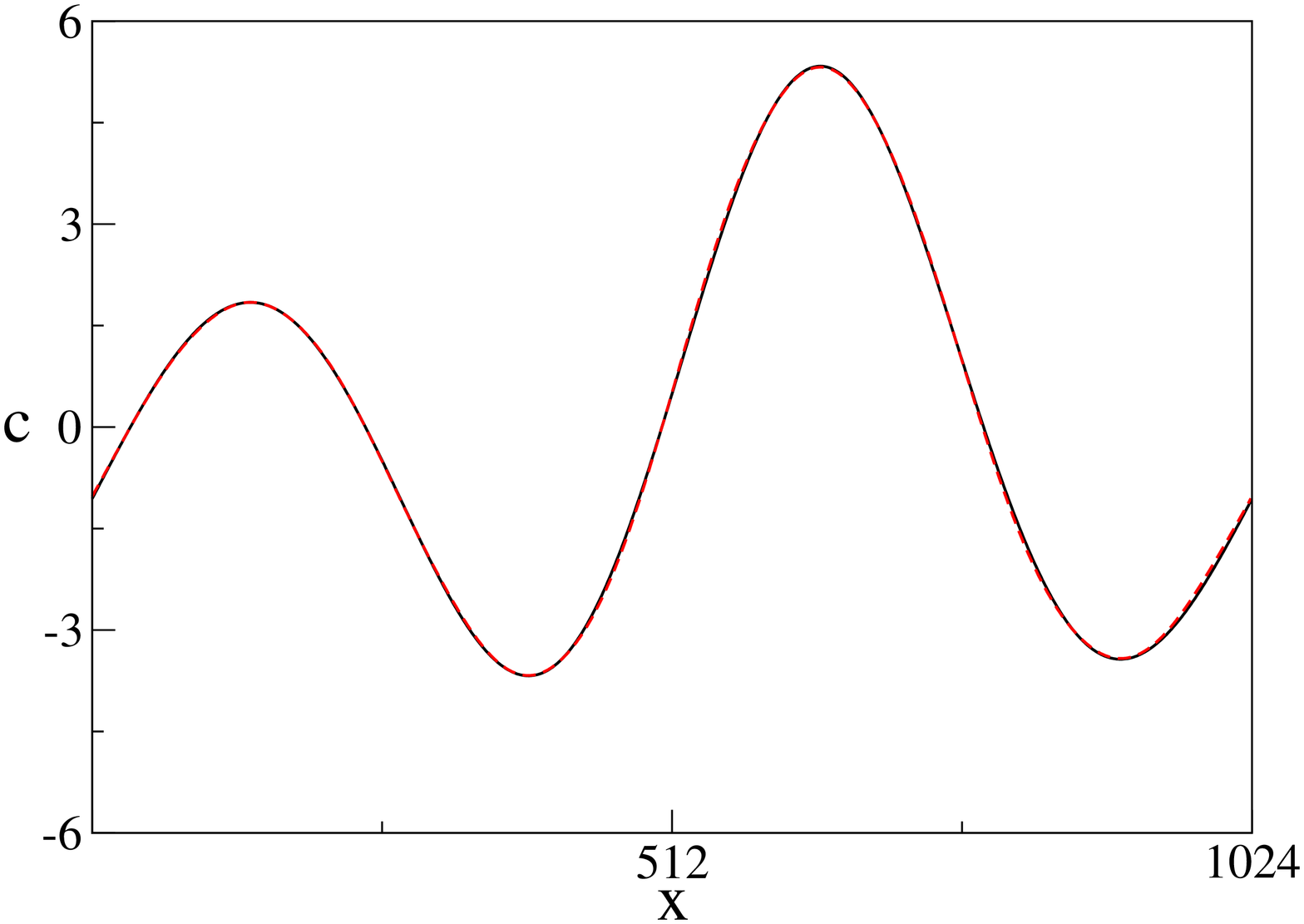}}}
\rotatebox{-90}{
\resizebox{6cm}{7cm}{
\includegraphics{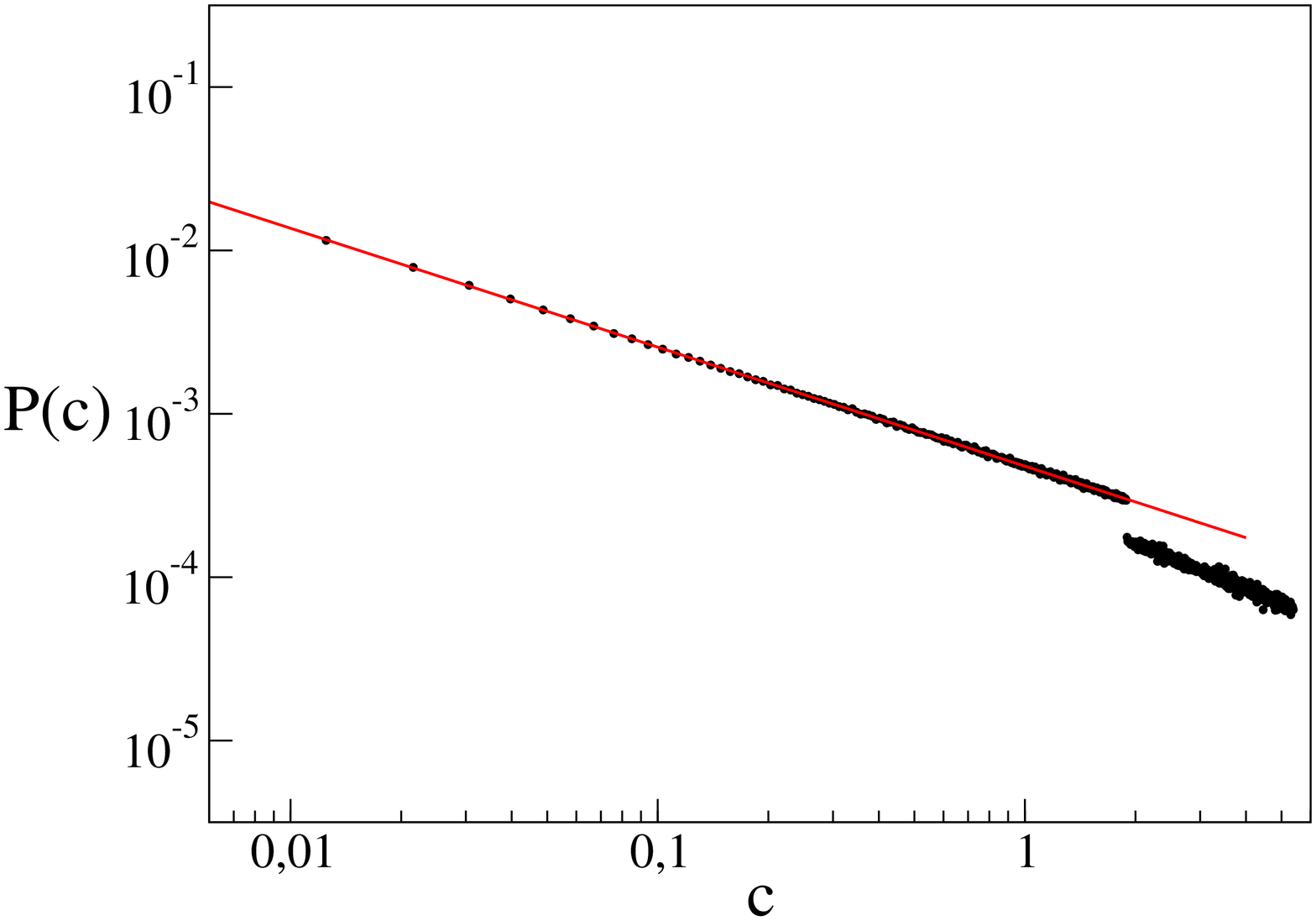}}}}
\caption{ (a) Left panel: Concentration fluctuations profile 
(solid line) obtained from the simulation data by a section plane cut through 
the hills and wells in Fig.2;  analytical function Eq.(\ref{a2}) (dashed line): 
the convex and concave curves are connected by summation of four
$q$-Gaussians;  (b) Right panel~: Concentration fluctuations
distribution (dots = data) with power law 
(solid line): $P(c) = A |c| ^{-q}$ with $q = 0.73$.}
\end{center}  
\label{Fig_C_profile}
\end{figure} 

Processes stemming from the generalized diffusion equation 
\cite{lutsko-boon} (or from the generalized Fokker-Planck equation 
provided some conditions are met for the drift and dissipative functions \cite{bukmann,borland,malacarne})
can produce solutions of the $q$-exponential type. 
In the present simulations, we observe that the concentration field fluctuations
exhibit a $q$-Gaussian profile -- while the system is still in the diffusive regime
($\sim t ^{1/2}$) -- which suggests that the concentration
$c(r,t)$ is governed by the generalized $d$-dimensional non-linear diffusion 
equation~\cite{lutsko-boon}~\footnote{On the r.h.s. of (\ref{a1}) one can 
rewrite formally $D\, \partial_r c^{\alpha}$ as $D_{\alpha} \,\partial_r c\,$,
where $D_{\alpha}\,=\,\alpha \,D \,c^{\,\alpha -1}$.}
\beqa{a1}
 \frac{\partial c(r,t)}{\partial t}  =
  \frac{1}{r^{d-1}} \frac{\partial}{\partial r} 
\left[ r^{d-1} D(r)\, \frac{\partial}{\partial r} c^{\alpha} (r,t) \right]\,.
\label{PME}
\eeqa 
A general solution to Eq.(\ref{a1}) is indeed a $q$-exponential  of the form
\cite{lutsko-boon}
\beq{a2}
c(r) = c_0\, e_q^{-\gamma_q \, |r|^{2}} 
\equiv c_0 \,[1-(1-q)\gamma_q \, |r|^{2}]^{\frac{1}{1-q}}\,,
\label{q-G-C_x}
\eeq 
where $c_0$ and $\gamma_q$ are time-dependent quantities, and  
$q+\alpha = 2$. This solution (taken at fixed~$t$) is used to perform a fit
to the simulation data, and the concentration profile found from the simulation 
results has exactly the form of (\ref{a2}) as shown in Fig.3a.

On the other hand, the velocity field is related to the concentration through
Darcy's law (see e.g. \cite{homsy}) which can be written for the stream function
$\psi$ (in two dimensions) as
\beq{a3}
 \left( \kappa_x \, \frac{\partial}{\partial x} +
 \kappa_y \, \frac{\partial}{\partial y} \right) \psi (x,y) =
 \frac{1}{R(x,y)} \, \nabla^2 \psi (x,y) \,,
\label{Darcy3}
\eeq 
with $ \kappa_x = {\partial c}/{\partial x}$, and 
$R = {\partial \ln \beta}/{\partial c}$, where $\beta$ is the damping function which, in the fingering simulations, controls the drag and depends on the concentration $c$.
Equation (\ref{a3}), coupled with the classical advection-diffusion equation for $c$, 
was solved numerically under conditions similar to those used
for the lattice Boltzmann simulations. For both the concentration field and the
stream function, we obtain $q$-Gaussians in agreement with the data obtained from the LBGK simulation results; in particular, for the stream function, we have
\beq{a4}
\psi (r) = \psi_0\, [1-(1-q')\phi_{q'} \, r^2]^{\frac{1}{1-q'}}\,,
\eeq 
where $\psi_0$ is a normalization constant

\begin{figure}
\begin{center}
\makebox{
\rotatebox{-90}{
\resizebox{6cm}{7cm}{
\includegraphics{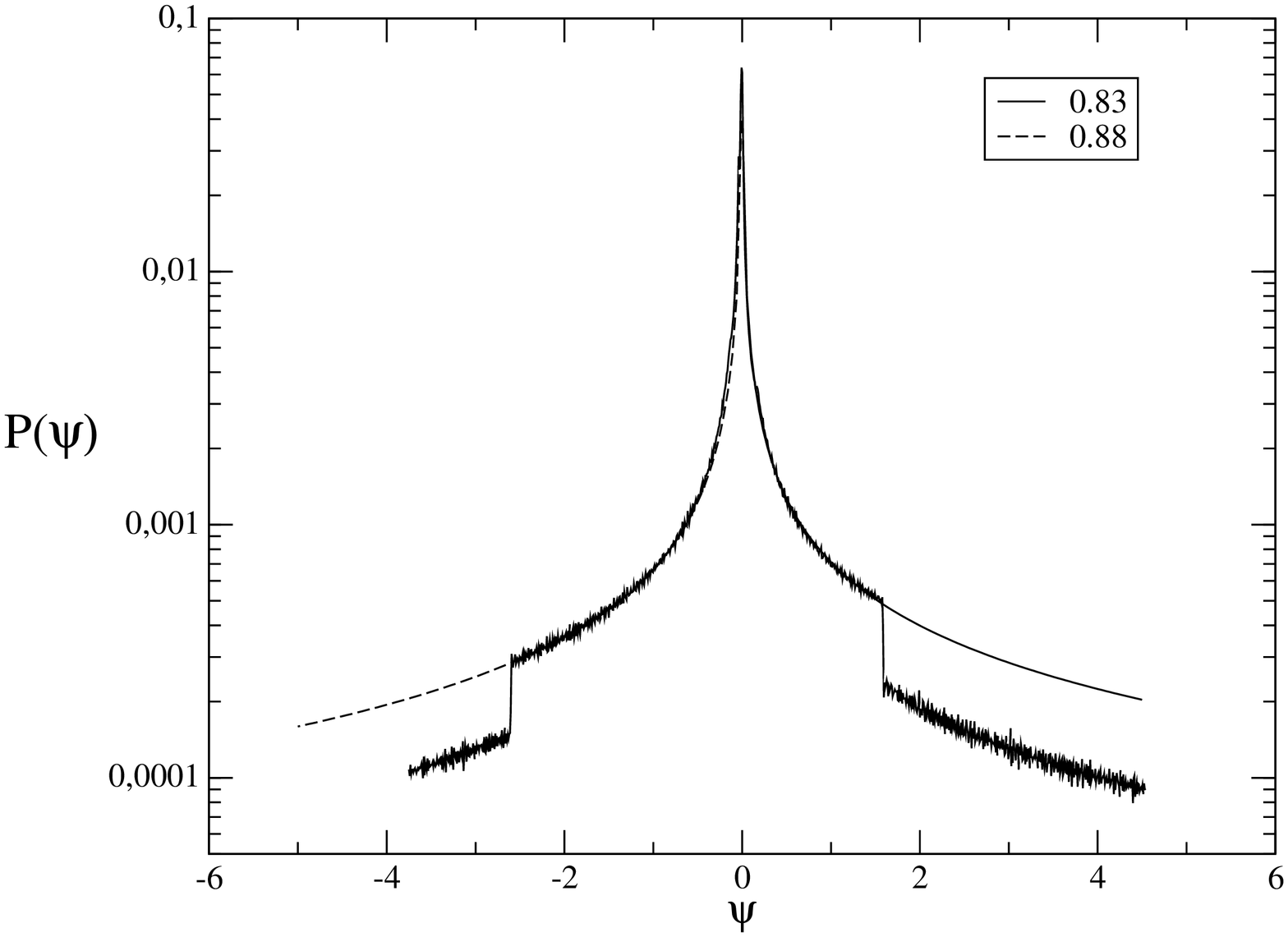}}}
\rotatebox{-90}{
\resizebox{6cm}{7cm}{
\includegraphics{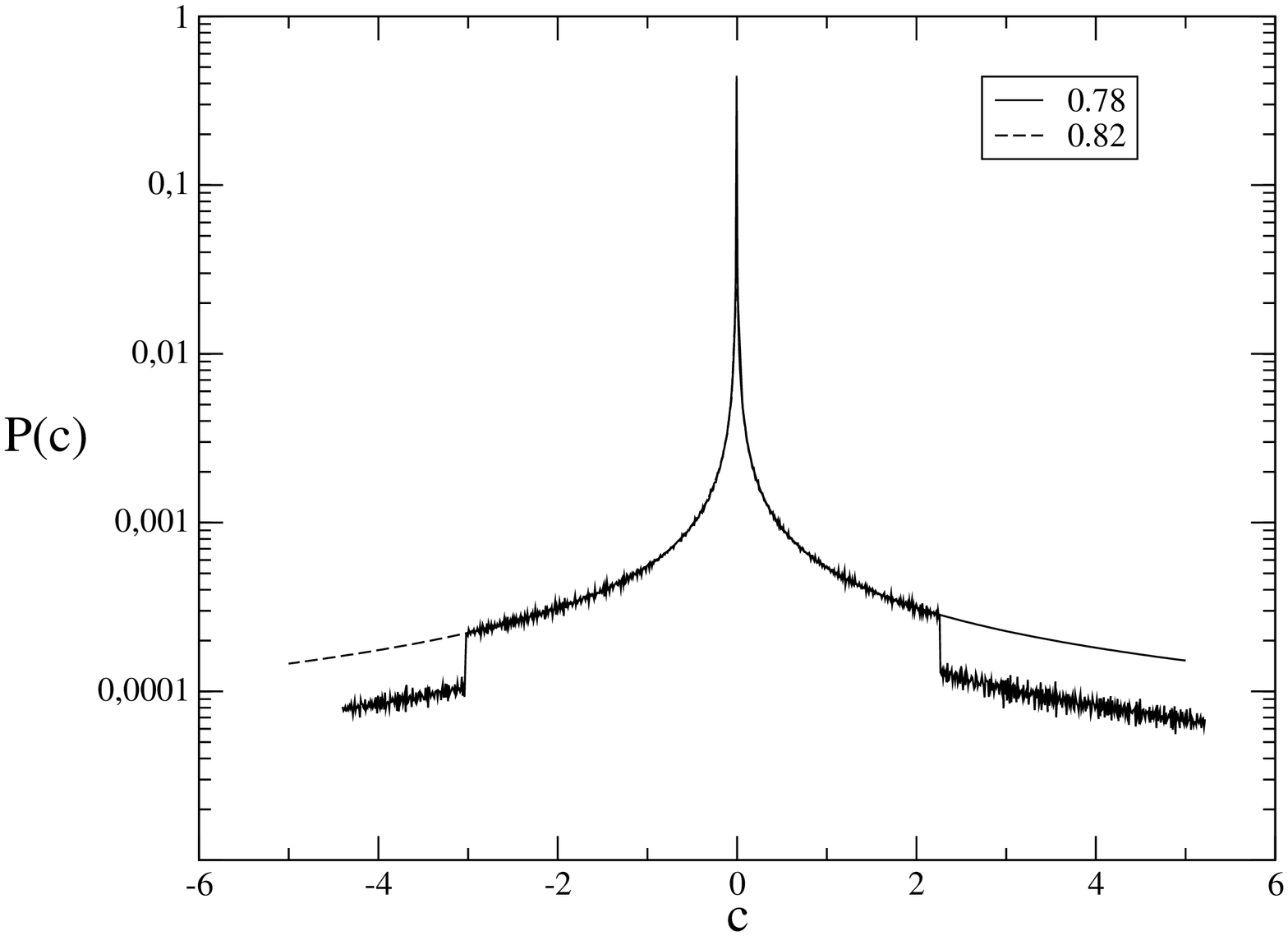}}}}
\caption{(a) Left panel: Stream function distribution (dots = data from numerical
solution of Eq.(\ref{a3})) with power law (solid and dashed lines) 
$P(\psi) \propto |\psi |^{-q'}$ (\ref{a10}) with 
$q' = 0.83$ for $|\psi > 0|$,  and $q' = 0.88$ for $|\psi < 0|$.
(b) Right panel: concentration field distribution (dots = data from numerical
solution of classical advection-diffusion equation coupled to Eq.(\ref{a3})) 
with power law (solid and dashed lines) $P(c) \propto |c|^{-q}$  with 
$q = 0.78$ for $|c > 0|$,  and $q = 0.82$ for $|c < 0|$.}
\label{Fig_psi_conc}
\end{center}
\end{figure}

We now show that the distribution which follows from the $q$-Gaussian 
spatial profile is a power law. Generalizing the $q$-exponential in  \Eq{a4} to an 
arbitrary power $\lambda$ of the radial variable~$r$, we have
 \beq{a5}
\psi (r) \,=\,e_q^{-\phi_{q'} \, r^\lambda}\,\equiv\, 
[1-(1-q') \phi_{q'} r^\lambda]^{\frac{1}{1-q'}}\,.
\label{q-G-v_r}
\eeq 
(for simplicity we have incorporated the normalization into $\psi $), and the
corresponding probability distribution function $P(\psi)$ reads (in $d$ dimensions)
\beq{a6}
P(\psi)\,=\, \varpi_d \, \int_0^\infty r^{d-1} |dr|  \,\delta(\psi (r)- \psi)\,,
\label{pdf_v}
\eeq
where $\varpi_d$ is an integration constant. Then (\ref{q-G-v_r}) is inverted to give 
$r^{\lambda}= \frac{1-\psi^{1-q'}}{\phi_{q'} (1-q')}
=-\frac{1}{\phi_{q'}}\,\ln_{q'} \psi$, which is used in (\ref{pdf_v}) to obtain
\beq{a7}
P(\psi) \,=\, \varpi_d \,  \frac{|r|^{d-\lambda} \psi^{-q'}} {\lambda \,\phi_{q'}}
       = \frac{\varpi_d}{\lambda \,\phi_{q'}^{d/\lambda}}
       \left[- \ln_{q'} \,\psi \right] ^\frac{d-\lambda}{\lambda} \psi^{-q'}\,.
\label{pdf_v_q}
\eeq 
For two-dimensional systems ($d=2$) with $q$-Gaussian spatial profile
($\lambda = 2$), we obtain the power law 
\beq{a10}
P(\psi)\,= \,\frac{\varpi_2}{2 \phi_{q'}} \, \psi^{-q'}\,,
\label{pdf_power}
\eeq 
which is exactly what we find for the stream function distribution shown
in Fig.4a  where a fit to the simulation data gives an exponent  $q < 1$ 
(indicating statistical importance of large values of the stream).
The distribution $P(c)$ (Fig.4b) is obtained along the same lines and yields the 
value $q \simeq 0.80 \pm 0.05$ to be compared with the value 
$0.73 \pm 0.05$ obtained from the lattice Boltzmann simulations (see Fig.3).
Singularities (elliptic singularities \cite{zaslavsky}) in the distribution function 
occur where the field has extrema (corresponding to the maximum of the absolute 
value of the blobs). According to Eq.(\ref{a1}), the value of
$q$ obtained from the concentration profile and distribution data 
yields $\alpha \simeq 1.25$ in the diffusion term of Eq.(\ref{a1}), 
thus an effective diffusion coefficient $D_{\alpha} \propto c^{0.25}$ 
(see footnote (2)). These results indicate that the generalized 
diffusion equation governs {\em effective} transport in fingering phenomena.

\begin{figure}
\begin{center}
\makebox{
\rotatebox{-90}{
\resizebox{6cm}{7cm}{
\includegraphics{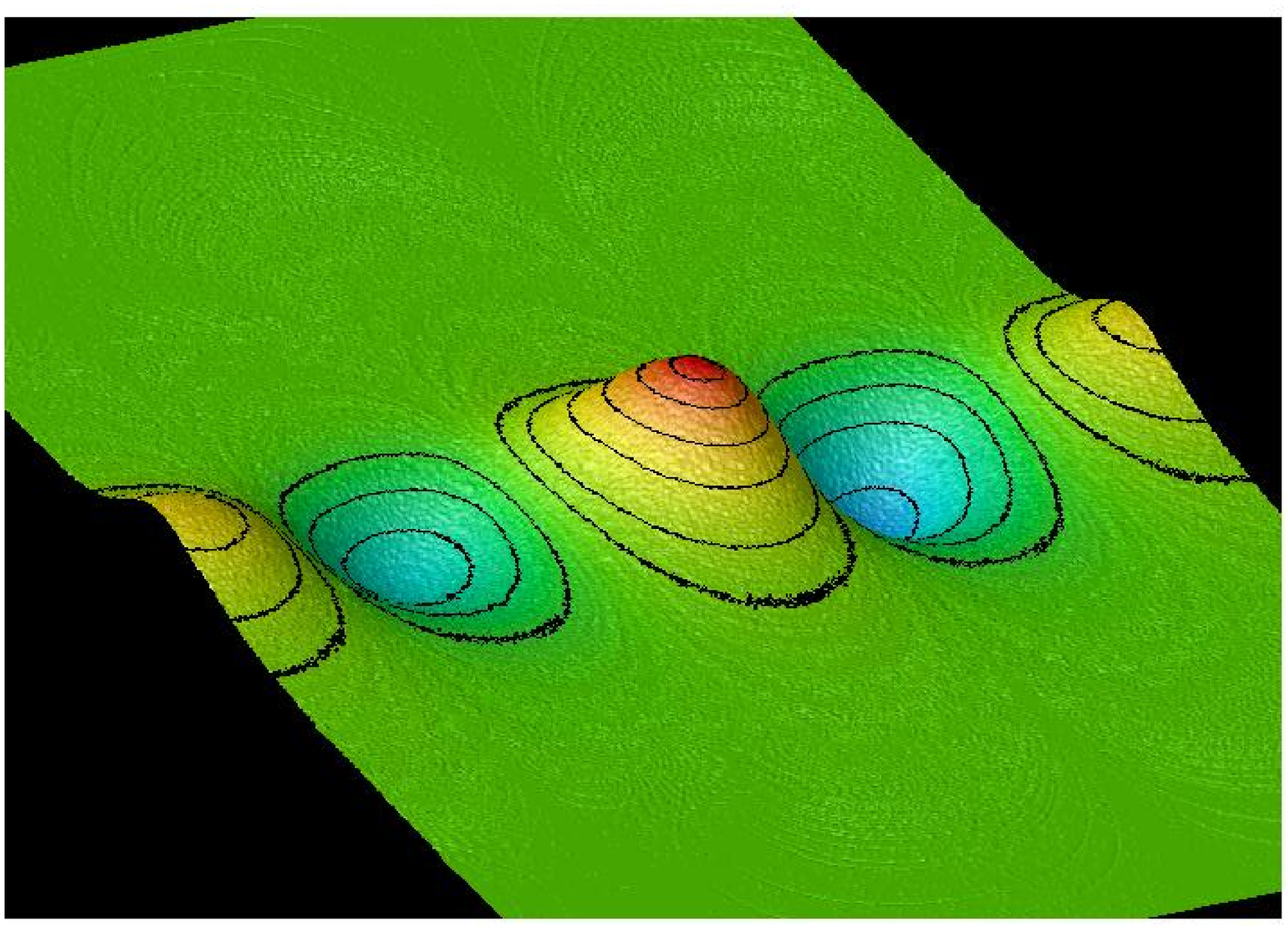}}}
\rotatebox{-180}{
\resizebox{7cm}{6cm}{ 
\includegraphics{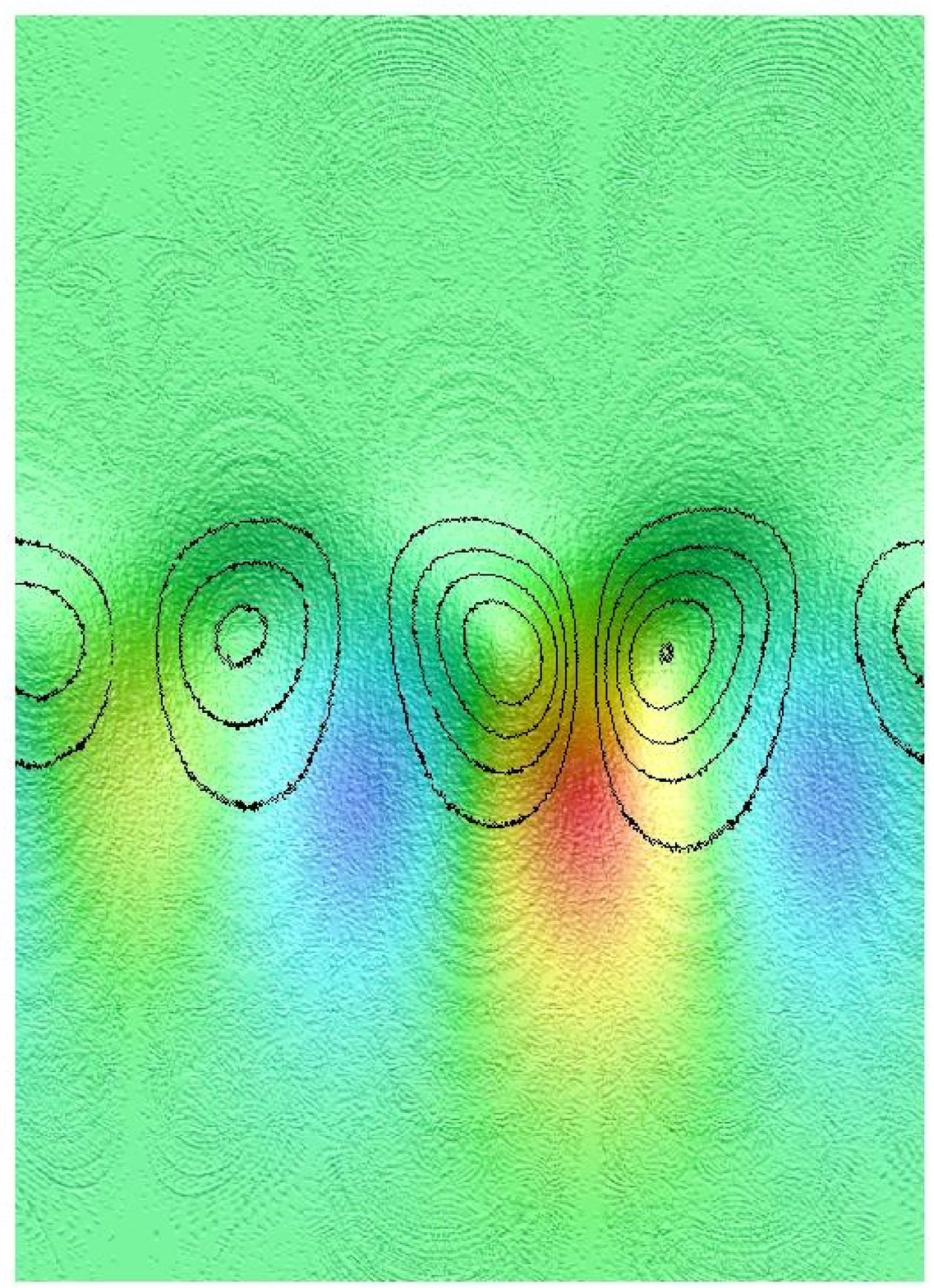}}}}
\caption{(a) Left panel:  Vorticity field showing landscape of 
$q$-Gaussian ``hills and wells" (flow direction South-East); 
simulation data with vorticity contour lines,
color code indicating highest positive values (red) to largest 
negative values (blue) of $\omega$. 
(b) Right panel: Contour lines of vorticity field (simulation data) superimposed 
on color concentration fluctuations field (simulation data); color code 
indicates highest positive values (red) to largest negative values (blue) 
of $c$. The inflection line of the concentration field ($C\simeq 0.5$) is 
about two thirds down from top of picture (flow direction downwards).}
\end{center}
\label{Fig_vort_conc}
\end{figure}

We also measured the vorticity field which, as shown in Fig.5a, exhibits a pattern 
of alternating vortices  which can be viewed as a landscape of hills and wells of 
radial $q$-Gaussians distributed along the concentration profile of the interfacial 
zone. Figure 5b shows the respective locations of the vorticity field (contour lines)
and concentration field (color coded). 

 Two questions arise from the present results: 
 (i) what is the physical picture of the {\em precursors} dynamics? and
 (ii) given that many systems exhibiting similar features have been analyzed
 with the nonextensive approach \cite{boon-tsallis}, we pose the reciprocal
 question: can one suggest a physical interpretation of {\em nonextensivity} 
 from the structure of the precursors~?
  
 \noindent (i) We observe that, before any fingering pattern becomes visible,
 the driving force triggers pre-transitional fluctuations in the concentration 
 field in the form of blobs organized spatially according to the wavelength of  
 the forthcoming fingers (Figs.2 and 3a). Concentration fluctuations
 modify locally the $C$-dependent drag coefficient thereby inducing
 shear. As a consequence, the initially uniform velocity field is 
 perturbed, and the resulting velocity fluctuations produce vortices 
 with alternating parities and whose spatial sequence matches 
 (with phase shift) the concentration fluctuation structures (Fig.5).
 The configuration shown in Fig.5b is reminiscent of the fingering 
 vortices depicted in a schematic in \cite{dewit} indicating that the 
 mechanism described in \cite{dewit} is already present in the precursors.
 
\noindent (ii) The system studied here is representative of a generic class 
of driven nonequilibrium systems with spatial fluctuations, where $q$-exponentials 
and power law distributions are the signature of long-range interactions,
and whose dynamical behavior is governed by advection-diffusion equations. 
What we have shown is that during the onset of fingering, one can identify 
precursors which exhibit features with mathematical properties typically 
encountered in nonextensive statistics. This similarity suggests a possible
interpretation of the mechanism of {\em nonextensivity}. 
The driving force produces a spatial sequence of alternating structures, which, 
if they were independent, would exhibit ordinary Gaussian profile originating 
from local diffusion centers ($\delta$-functions), and would be described by classical diffusion. However, when growing, these Gaussians overlap,  and what the analysis shows is  that the structure formed by these overlapping Gaussians can be cast into 
a {\em sum} of scale invariant $q$-Gaussians with finite support ($q < 1$), 
i.e. these $q$-Gaussians are independent as corroborated by the fact that the 
power law distribution is shown (see Eq.(\ref{a10})) to follow from space 
integration of a {\em single} $q$-exponential. 

{This work was supported by a grant from the 
{\em European Space Agency} and {\em PRODEX} (Belgium) 
under contract ESA/14556/00/NL/SFe(IC).}

\end{document}